\documentclass[twoside]{article}
\usepackage{fleqn,espcrc2}
\def\beq{\begin{equation}}
\def\eeq{\end{equation}}
\def\bea{\begin{eqnarray}}
\def\eea{\end{eqnarray}}
\def\ba{\begin{array}}
\def\ea{\end{array}}
\def\del{\partial}

\def\t{\widetilde}
\def\G{\Gamma}
\def\O{\Omega}
\def\L{\Lambda}
\def\e{\epsilon}
\def\a{\alpha}
\def\sC{\not{\hbox{\kern-2.5pt {\rm C}}}}
\def\sY{\not{\hbox{\kern-1.5pt {\rm Y}}}}
\def\sV{\not{\hbox{\kern-1.5pt {\rm V}}}}
\def\sF{\not{\hbox{\kern-1.5pt {\rm F}}}}
\def\osV{\not{\hbox{\kern-1.5pt {$\overline{\rm V}$}}}}

\newcommand{\AmS}{{\protect\the\textfont2
  A\kern-.1667em\lower.5ex\hbox{M}\kern-.125emS}}
\title{Supersymmetry and the Systematics of T-duality Rotations in
Type-II Superstring Theories}
\author{S. F. Hassan
\thanks{\tt fawad.hassan@helsinki.fi}
\thanks{Contribution to the proceedings of the D. V. Volkov memorial
conference on ``Supersymmetry and Quantum Field Theory'', Kharkov,
July 25-29, 2000 (to appear in the Nucl. Phys. B Conference
Supplements).}
\\[.1cm]
Helsinki Institute of Physics,
P.O. Box 64, FIN-00014, University of Helsinki, Finland}
\begin{document}

\begin{abstract}
We describe a systematic method of studying the action of the
T-duality group $O(d,d)$ on space-time fermions and R-R field
strengths and potentials in type-II string theories, based on
space-time supersymmetry. The formalism is then used to show that the
couplings of non-Abelian D-brane charges to R-R potentials can be
described by an appropriate Clifford multiplication.
\\[.2cm]
{\it Report No: HIP-2001-07/TH, hep-th/0103149}
\vspace{-.5cm}
\end{abstract}
\maketitle
\section{Introduction:}
Let us denote the massless NS-NS background fields by $G_{MN}, B_{MN}$
and $\phi$, and the R-R {\it n}-form field strengths by
$F^{(n)}_{M_1\cdots M_n}$, where $n$ is $even$ in type-IIA and $odd$
in type-IIB theory. It is customary to define two sets of R-R
potentials, which we denote by $C^{(n)}_{M_1\cdots M_n}$ and
$C'^{(n)}_{M_1\cdots M_n}$. Denoting by $\mathbf F$ and $\mathbf C$
the sums of {\it n}-forms $\sum_{n=1}^{n=9}F^{(n)}$ and
$\sum_{n=0}^{n=8}C^{(n)}$ (similarly for $\mathbf C'$), the relations
among them can be written as
\beq
\mathbf F = d\mathbf C - H\wedge \mathbf C
          = {\rm e}^B \wedge d\mathbf C'
\label{F-C}
\eeq
Here, $H=dB$ and the two sets of potentials are related by
$\mathbf C'= \mathbf C\wedge {\rm e}^{-B}$. Also we denote the two
gravitinos, the two dilatinos and two supersymmetry transformation
parameters by $\Psi_{\pm M}$, $\lambda_\pm$ and $\e_\pm$,
respectively. In type-IIA theory the fermions labelled by $\pm$ have
opposite chiralities, while in type-IIB they have the same chirality.

Let $X^M, (M=0,\cdots 9)$ denote the space-time coordinates and
consider background field configurations which are independent of the
$d$ coordinates $X^i, (i=1,\cdots d)$, but may vary with the remaining
$10-d$ coordinates $X^\mu, \mu=0, d+1,\cdots 9$. The T-duality group
$O(d,d)$ acts on these backgrounds such that the equations of motion,
when restricted to $X^i$-independent fields, remain invariant. The
transformation of NS-NS fields under this group is well known and can
be obtained in a variety of ways. Here we describe a systematic
procedure, developed in \cite{H1,H2}, for obtaining the
transformations of R-R field strengths and potentials. As a
by-product, one also obtains the transformations of the gravitinos,
dilatinos and supersymmetry parameters. In the last section, we
describe an application of the results to the coupling of non-Abelian
D-brane charges to R-R backgrounds \cite{HM}.

The transformation of R-R potentials $C^{(n)}$ under a {\it single}
T-duality was first obtained in \cite{BHO} in the effective low-energy
theory. More recently, the case of single T-duality in the presence of
supersymmetry has also been considered in \cite{CLPS,KS,KR}.
In \cite{HT,W} it was observed that the potentials $C'^{(n)}$
transform in the spinor representations of the T-duality group, verified
directly in \cite{FOT}. Also, see \cite{G}. However, this cannot be
used to obtain the transformation of $C^{(n)}$ and $F^{(n)}$ under the
non-trivial elements of the full T-duality group.

\subsection{Supersymmetry Transformations}
We obtain the transformations of massless R-R and R-NS fields under {\it
non-trivial} $O(d,d)$ transformations, by demanding compatibility
between T-duality and space-time supersymmetry. In type-II theories,
this leads to the invariance of the equations of motion.

Not all terms in the supersymmetry transformations are needed for the
analysis. To see the basic structure, let us first consider type-II
superstring theories in flat space. The two space-time supersymmetry
transformations $\delta_\pm$ act independently on the left $(+)$ and
right $(-)$ moving sectors of the worldsheet by interchanging Ramond
states $|R_\pm\rangle$ with Neveu-Schwarz states $|NS_\pm\rangle$. For
example, schematically, we can write
\bea
\delta_-|NS_+,R_-\rangle&=&|NS_+,NS_-\rangle\e_-\,,\nonumber\\
\delta_+|R_+,NS_-\rangle&=&|NS_+,NS_-\rangle\e_+\,,\nonumber\\
\delta_-|R_+,NS_-\rangle&=&|R_+,R_-\rangle\e_-\,.\nonumber
\eea
If we know the action of the T-duality group on the NS-NS sector, then,
demanding compatibility with supersymmetry, the first two equations
determine the T-duality transformations of the R-NS sector and those of
$\e_\pm$. One can then use the third equation to determine the
T-duality transformation of the R-R sector.

In the presence of background fields, the supersymmetry
transformations are more complicated. However, terms with the lowest
powers of fermions still retain the flat-space structure. For example,
for the gravitinos $\Psi_{\pm M}$, corresponding to the above three
equations we have \cite{JHS,LR,H1}
\bea
\delta_- \Psi_{- M}&=&(\del_M+\frac{1}{4} W^-_{Mab}\G^{ab})\,\e_-
 + \cdots\,,
\label{susy1}\\
\delta_+ \Psi_{+ M} &=& (\del_M+\frac{1}{4} W^+_{Mab}\G^{ab})\,\e_+
 + \cdots\,,
\label{susy2}\\
\delta_-\Psi_{+ M}&=&\frac{1}{2(8)}e^{\phi}\sF\,\G_M\,\e_-+\cdots\,.
\label{susy3}
\eea
Here, $W^\pm_{Mab}=w^\pm_{Mab}\mp\frac{1}{2} H_{Mab}$ are the
torsionful spin-connections and $a, b$ are the 10-dimensional Lorentz
frame indices. $\sF$ is a bispinor constructed out the R-R field
strengths as
\beq
\sF= \sum_n \frac{(-1)^n}{n!} F^{(n)}_{M_1\cdots M_n}
\G^{M_1\cdots M_n}\,.
\label{check}
\eeq
The dots ``$\cdots$'' in equations (\ref{susy1})-(\ref{susy3})
represent terms containing cubic powers of the spinors. We do not
write these out explicitly but emphasize that their contribution will
be accounted for in the final results. The strategy now is to first
work out the transformation of $W^{\pm}_{Mab}$ under non-trivial
elements of the duality group. Then from equations (\ref{susy1}) and
(\ref{susy2}) we obtain the corresponding transformation of $\Psi_{\pm
M}$ and $\e_\pm$. In turn, these can be used in (\ref{susy3}) to find
the transformation of $\sF$. Let us first review the
transformation of the metric under the duality group.

\subsection{T-duality of the Metric}

Not all elements of the T-duality group $O(d,d)$ transform the
background fields non-trivially. In fact, $GL(d, R)$ transformations of
the coordinates $X^i$, as well as constant shifts in $B_{ij}$ are
subgroups of the T-duality group. These are the {\it trivial}
elements. The rest act non-trivially and are symmetries of the
equations of motion only when the fields do not depend on the $d$
coordinates $X^i$. Let us parameterize the $O(d)\times O(d)$ subgroup
of $O(d,d)$ by matrices ${\cal S}\in O(d)$ and ${\cal R}\in O(d)$. All
non-trivial elements fall in this subgroup and are parameterized by
${\cal S}$ and ${\cal R}$ with ${\cal S}\neq {\cal R}$. The case
${\cal S}= {\cal R} (={\cal O}, {\rm say})$ is equivalent to a
coordinate transformation $X^i\rightarrow {\cal O}^i_j X^j$ and
belongs to the $O(d)$ subgroup of $GL(d, R)$, already classified as
trivial. We will be interested in the transformation of fields under
the $O(d)\times O(d)$ subgroup parameterized by the $d$-dimensional
matrices ${\cal S},{\cal R}$. In order to write the transformation of
the 10-dimensional fields in a compact way, it is convenient to
enlarge these to 10-dimensional matrices $S$ and $R$ simply by adding
the identity matrix for the extra dimensions. By definition, these
satisfy $S^M_{\,K}\hat\eta^{KL}S_{L}^{\,\,\,N}=\hat\eta^{MN}$,
similarly for $R$. Here $\hat\eta$ is the flat metric with respect to
which the orthogonal groups are defined.

The transformation of the metric $G_{MN}$ under the $O(d)\times O(d)$
subgroup can be written in {\it two equivalent} ways \cite{H3} ,
\beq
\t G^{-1}= Q_- G^{-1}Q^T_- = Q_+ G^{-1}Q^T_+\,,
\label{Gpm}
\eeq
where the matrices $Q^M_{\pm N}$ are given by
\bea
Q_-=\frac{1}{2}\left[(S+R)+(S-R)\eta^{-1}(G+B)\right]\,,\\
Q_+=\frac{1}{2}\left[(S+R)-(S-R)\eta^{-1}(G-B)\right]\,.
\eea
From the structure of $S$ and $R$ one can see that
\beq
(Q_\pm)^\mu_{\,\,j}=0
\,,\quad (Q_\pm)^\mu_{\,\,\nu}=\delta^\mu_{\,\nu}\,,
\label{Qcompo}
\eeq
with similar equations for $Q^{-1}_\pm$. We do not write down the
transformations of other NS-NS fields.

\subsection{The Local Lorentz Twist}

To couple fermions to gravity one introduces vielbeins $e^M_a$ through
$G^{MN}=e^M_a\eta^{ab}e^N_b$. The observation that the T-duality
transformation of the metric can be written in two equivalent ways
means that there are two possible vielbeins in the dual theory,
\beq
\t e^M_{(-)a} = Q^M_{-N}e^N_a\,,\qquad
\t e^M_{(+)a} = Q^M_{+N}e^N_a\,.
\label{epm}
\eeq
They are related by a Lorentz transformation $\L^a_{~b}$,
\beq
\t e^M_{(+)b} = \t e^M_{(-)a}\L^a_{~b}\,,\quad
\L^a_{~b}=e^a_M (Q^{-1}_-Q_+)^M_{\,\,N}e^N_b\,.
\label{llt}
\eeq
In string theory, the vielbeins $e^a_M$ and the associated {\it local}
Lorentz frame can originate in either the {\it right-}, or the {\it
left-} moving sector of the worldsheet theory. Depending on this origin,
the vielbeins transform to either $\t e_-$ or $\t e_+$. Alternatively,
as implied by (\ref{llt}), we can choose a single set of vielbeins,
say, $\t e_{(-)}$, for both the right and left moving sectors, but
assume that the associated local Lorentz frames are now twisted with
respect to each other by an amount $\L^a_{~b}$. Below, we show that
this {\it local Lorentz twist}, induced by the action of the T-duality
group, can be undone by absorbing it into the Ramond sector. This, in
turn, dictates the transformation of the space-time spinors and R-R
fields.

\section{T-duality and Supersymmetry}
Now we have the ingredients needed to analyse the effect of T-duality
on the supersymmetry variations (\ref{susy1})-(\ref{susy3}). First we
need the transformation of the spin-connections $W^{\pm}_M$. In the
dual theory, we denote these by $\t W^{\pm}_{(-)M}$. The
subscript $(-)$ indicates that they are defined with respect to $\t
e_{(-)}$. One can show that \cite{H1,H2}
\bea
\t W^-_{(-)Mab}&=&W^-_{Nab}(Q^{-1}_+)^N_{~M}\,,\label{W--}\\
\t W^+_{(+)Mab}&=&W^+_{Nab}(Q^{-1}_-)^N_{~M}\,.\label{W++}
\eea
$\t W^+_{(-)M}$ can now be easily obtained from $\t W^+_{(+)M}$ by
noting that $\t e_{(+)}$ and $\t e_{(-)}$ are related by a Lorentz
transformation $\L$ (\ref{llt}). Let $\O$ denote the spinor
representation of $\L$ defined by
\beq
\O^{-1}\G^a\O = \L^a_{~b}\G^b\,.
\label{spin-llt}
\eeq
Then, using (\ref{Qcompo})-(\ref{spin-llt}), one sees that the duality
transformations of the spinorial covariant derivatives,
$D^\pm_M = \del_M +\frac{1}{4}W^\pm_{Mab}\G^{ab}$ are given by
\beq
\t D^-_{(-)M}=(Q^{-1}_+)^N_{~M} D^-_N \,,
\eeq
\beq
\t D^+_{(-)M}=\O\t D^+_{(+)M}\O^{-1}=(Q^{-1}_-)^N_{~M}
\O D^+_N\O^{-1}\,.
\eeq

\subsection{Transformation of Spinors}
Using the transformation of the spinorial covariant derivatives
above, in the supersymmetry variations (\ref{susy1}) and
(\ref{susy2}), one obtains the T-duality action on the supersymmetry
parameters $\e_\pm$ as
\beq
\t\e_-=\e_-\,,\qquad \t\e_+=\O\e_+\,.
\label{tdepm}
\eeq
As for $\Psi_{\pm M}$, it seems that their transformation is
determined only up to terms cubic in the spinors. There are two
sources of 3-spinor corrections: {\it i)} equations
(\ref{susy1}) and (\ref{susy2}) contain 3-spinor terms that we have
not taken into account, {\it ii)} the T-duality transformations
$\Psi_{\pm M}$ and $\delta_\pm\Psi_{\pm M}$ differ by 3-spinor terms
since $\delta_\pm Q_\mp$ and $\delta_\pm \O$ are both quadratic in the
spinors. It turns out that contributions of type {\it i} and {\it ii}
cancel each other and the T-duality action on the gravitinos is given
by
\bea
\t\Psi_{-M}&=&\Psi_{-N}(Q^{-1}_+)^N_{~M}\,,
\label{tdpsi-}\\
\t\Psi_{+M}&=&\O\Psi_{+N}(Q^{-1}_-)^N_{~M}\,.
\label{tdpsi+}
\eea
The absence of 3-spinor corrections can be confirmed by verifying that
they are not allowed by the supersymmetry variations of the NS-NS
fields. This means that T-duality does not mix terms with different
space-time fermion numbers.

A similar analysis of the supersymmetry variations of the dilatinos
gives
\bea
\t\lambda_- \!\!\!&=&\!\!\! \lambda_-
-\frac{1}{2} \G_i \left[ Q^{-1}_-(S-R)\hat\eta^{-1}\right]^{ij}
\Psi_{-j}\,,\nonumber\\
\t\lambda_+ \!\!\!&=&\!\!\!  \O\Big(\lambda_+
+\frac{1}{2} \G_i \left[ Q^{-1}_+(S-R)\hat\eta^{-1}\right]^{ij}
\Psi_{+j} \Big)\,.\nonumber
\eea

\subsection{The R-R Field Strength Bispinor}
In the $O(d,d)$ transformed theory, the supersymmetry variation
(\ref{susy3}) becomes
\beq
\delta_-\t\Psi_{+M}=\frac{1}{2(8)}\,e^{\t\phi}\, \t{\sF}\,\t\G_{(-)M}
\,\t\e_-\,\,+\,\cdots\,.
\label{dtG-RR}
\eeq
Then, using (\ref{tdepm}) and (\ref{tdpsi+}) along with the dilaton
transformation, $e^{\phi-\t\phi}=\sqrt{det Q_-}$, one obtains
\beq
\t{\sF}=
\sqrt{\,\det\,Q_-}\,\O\,\sF\,,
\label{tdcalF}
\eeq
where now,
\beq
\t{\sF}=\sum_n\frac{(-1)^n}{n!}\,\t F^{(n)}_{N_1\cdots N_n}\,
\t\G^{M_1\cdots M_n}_{(-)}\,,
\label{tbispin}
\eeq
with $\t\G^M_{(-)}= Q^M_{-N}\G^N$. To find the transformation of
the component field strengths $F^{(n)}$, as well as those of the
associated R-R potentials $C^{(n)}$ and $C'^{(n)}$, we need the
explicit form of the spinor representation $\O$ of the Lorentz twist
$\L$.

\section{Transformations of R-R Fields}
The T-duality group $O(d,d)$ has elements with determinant $+1$ and
$-1$. A representative of the latter class is a {\it single} T-duality
with respect to a coordinate, say $X^9$. This is a ${\mathbf Z_2}$
transformation which corresponds to the choice $S=1$,
$R^M_N=\delta^M_N-2\delta^M_9\delta^9_N$. It interchanges type-IIA and
type-IIB theories \cite{DHS}. The elements with determinant $+1$, of
course, form the group $SO(d,d)$ which acts within type-IIA or
type-IIB theory. We consider these two cases separately.
\subsection{The Single T-duality Case}
For a single T-duality, say, along $X^9$, the spinor representation
$\O$ is given by \cite{H1}
\beq
\O = a \sqrt{G^{-1}_{99}}\G_{11}\G_9\,.
\label{td1-spinor}
\eeq
Here, $a$ is an arbitrary sign that we choose to be $+1$ in going from
IIA to IIB and $-1$ {\it vice versa}, so that the transformation
squares to $+1$. Using this in (\ref{tdcalF}), one obtains
\bea
\t F^{(n)}_{9i_2\cdots i_n}\!=\!- F^{(n-1)}_{i_2\cdots i_n}
+(n\!-\!1) G^{-1}_{99} G_{9[i_2} F^{(n-1)}_{9i_3\cdots i_n]}\,,
\!\!\!\!&&\nonumber
\eea
\beq
\t F^{(n)}_{i_1i_2\cdots i_n}\!=\!-F^{(n+1)}_{9i_1\cdots i_n}
-n B_{9[i_1} \t F^{(n)}_{9i_2\cdots i_n]}\,.
\label{td1-F}
\eeq
Relation (\ref{F-C}) between the $F^{(n)}$ and $C^{(n)}$ then gives
\bea
\t C^{(n)}_{9i_2\cdots i_n}\!\!=\! C^{(n-1)}_{i_2\cdots i_n}
-(n\!-\!1) G^{-1}_{99} G_{9[i_2} C^{(n-1)}_{9i_3\cdots i_n]}\,,
\nonumber
\eea
\beq
\t C^{(n)}_{i_1i_2\cdots i_n}\!\!=\! C^{(n+1)}_{9i_1\cdots i_n}
-n B_{9[i_1} \t C^{(n)}_{9i_2\cdots i_n]}\,,
\label{td1-C}
\eeq
as first obtained in \cite{BHO}. The transformation of $C'^{(n)}$
can be obtained using $\mathbf C\;'=\mathbf C\wedge {\rm e}^{-B}$, as
\beq
\t C'^{(n)}_{9i_2\cdots i_n}=C'^{(n-1)}_{i_2\cdots i_n}\,,\qquad
\t C'^{(n)}_{i_1i_2\cdots i_n}=C'^{(n+1)}_{9i_1\cdots i_n}\,.
\label{td1-C'}
\eeq
Now we make three simple observations that will greatly simplify
things in the next subsection:
\begin{itemize}
\item Under a {\it single} T-duality, $F^{(n)}$ and $C^{(n)}$
transform in the same way, up to a sign. The sign difference
disappears when considering even number of single T-dualities.
\item The transformation of $C'^{(n)}$ is independent of $G_{MN}$
and $B_{MN}$. Hence, it transforms in the same way as $C^{(n)}$ would
in the flat-space $G_{MN}\rightarrow\hat\eta_{MN}$,
$B_{MN}\rightarrow 0$.
\item Being $n$-forms, $F^{(n)}$, $C^{(n)}$ and $C'^{(n)}$ transform
in the same way under $GL(d,R)$ transformations of the coordinates
$X^i$, in particular, under its $SO(d)$ subgroup.
\end{itemize}

\subsection{The $SO(d,d)$ Case}
The spinor representation of the Lorentz twist induced by an $SO(d,d)$
transformation of the backgrounds is given by \cite{H2}
\beq
\O=2^{-\frac{d}{2}}
\sqrt{\frac{\det({\cal Q}_- + {\cal Q}_+)}{\det{\cal Q}_-}}\,
\AE
(-\frac{1}{2}{\cal A}^{ij}\G_{ij})\,.
\label{sodd-spinor}
\eeq
Here, ${\cal Q}_\pm$ stand for the $d\times d$ blocks of $Q_\pm$
spanned by the index $i$, and $\AE$ stands for an exponential-like
expansion with the products of all $\G$-matrices antisymmetrized.
The ${\cal A}^{ij}$ appearing above is given by
\beq
{\cal A}^{ij}\!\!=\!\!\left[\hat\eta_d({\bf 1}_d-{\cal S}^{-1}{\cal R})^{-1}
({\bf 1}_d +{\cal S}^{-1}{\cal R})\!+\!{\cal B}\right]^{-1}_{ij}.
\label{Aij}
\eeq
Substituting (\ref{sodd-spinor}) in (\ref{tdcalF}) one obtains
\bea
\t F_{M_1\cdots M_n}\hspace{-.2cm}&=&\hspace{-.2cm}
2^{-\frac{d}{2}}\sqrt{\det({\cal Q}_- + {\cal Q}_+)}\,
\sum_{p=0}^{[d/2]}\sum_{r=0}^{2p}
\nonumber\\
&&\hspace{-1.8cm}
\times\Big[\frac{(-1)^p}{p!
2^p}\,\frac{(2p)!}{(2p-r)!}\, {}^nC_r\,
{\cal A}^{[i_1i_2}\cdots {\cal A}^{i_{2p-1}i_{2p}]}
\nonumber\\
&&\hspace{-.5cm}
G_{i_1N_1}\cdots G_{i_rN_r}F_{i_{2p}\cdots i_{r+1}N_{r+1}
\cdots N_n}\,\Big]
\nonumber\\
&&\qquad\times
(Q^{-1}_-)^{N_1}_{\,\,[M_1}\cdots (Q^{-1}_-)^{N_n}_{\,\,M_n]}\,,
\label{SOdd-F}
\eea
where, ${}^nC_r$ are the binomial expansion coefficients.

\subsection{Transformation of Potentials $C^{(n)}$}
Unlike the case of a single T-duality, the transformation of $C^{(n)}$
and $C'^{(n)}$ cannot be easily worked out by using equation (\ref{F-C}).
However, these can be obtained by a simple construction combined with
the three observations made in subsection 3.1.

Any non-trivial $SO(d,d)$ transformation can, in principle, be
constructed as a combination of single T-duality transformations and
appropriately chosen coordinate rotations. To see this, let
$\hat r$ denote a unit vector in $d$ dimensions. Any $O(d)$ rotation,
say ${\cal R}$, can be decomposed as a product of reflections
$T_{r_k}$ about planes perpendicular to properly chosen axes $\hat
r_k$, {\it i.e.}, ${\cal R}=T_{r_n}\cdots T_{r_1}$. Equivalently, it
can be written as a product of reflections $T_i$ about planes
perpendicular to the coordinate axes $\hat x^i$, and properly chosen
rotations ${\cal O}_k$ that rotate the coordinate axes $\hat x^i$ into
the reflection axes $\hat r_k$. Also setting ${\cal S}$ to a product
of the rotations ${\cal O}_k$, we get
\beq
{\cal R}=T_{i_n}{\cal O}_{k_n}\cdots T_{i_1}{\cal O}_{k_1}\,,\quad
{\cal S}= {\cal O}_{k_n}\cdots {\cal O}_{k_1}\,.
\label{TO}
\eeq
Then the non-trivial $O(d,d)$ transformation implemented by ${\cal R}$
and ${\cal S}$ corresponds to a sequence of {\it single} T-duality
transformations with ${\cal R}_i=T_{i}$, ${\cal S}_i=1$, intertwined
with coordinate rotations ${\cal R}_k={\cal S}_k={\cal O}_k$.

This construction implies that the non-trivial $SO(d,d)$
transformations of $F^{(n)}$ in (\ref{SOdd-F}) can, at least in
principle, be constructed by applying a succession of {\it single}
T-dualities (\ref{td1-F}) and coordinate rotations, using the
decomposition (\ref{TO}).

As emphasized in subsection 3.1, $F^{(n)}$ and $C^{(n)}$ transform in
the same way under a single T-duality (the sign difference is
immaterial for even number of T-dualities). They also transform in the
same way under coordinate rotations. Therefore, if we
construct the action of an $SO(d,d)$ element on $C^{(n)}$, by using its
decomposition in terms of single T-dualities and rotations, then we will
end up with the  same equation as for $F^{(n)}$, {\it
i.e.,}(\ref{SOdd-F}), but now with $F^{(n)}$ replaced by $C^{(n)}$. In
terms a bispinor ${\sC}$, defined analogously to (\ref{check}), the
transformation is
\beq
\t{\sC} = \sqrt{Q_-}\Omega\,\sC\,.
\label{Odd-C}
\eeq
Needless to say, this formula is also valid for non-trivial $O(d,d)$
transformations.

\subsection{Transformation of Potentials $C'^{(n)}$}
As observed in subsection 3.1, a {\it single} T-duality acts on
$C'^{(n)}$ in the same way as it would act on $C^{(n)}$ in a flat
space with $B_{MN}\rightarrow 0$ and $G_{MN} \rightarrow
\hat\eta_{MN}$. Furthermore, both potentials transform in the same way
under coordinate rotations. The decomposition (\ref{TO}) then implies
that non-trivial $O(d,d)$ transformations of $C'^{(n)}$ are given by
the same formula as that for $C^{(n)}$ with $B_{MN}=0$ and $G_{MN}$
set to $ \hat\eta_{MN}$.

This result can also be written in a compact form similar to equation
(\ref{Odd-C}). For this, let us define a set of $SO(1,9)$ Gamma
matrices $\hat\G^M$ by
\beq
\{\hat\G^M,\hat\G^N\}=\hat\eta^{MN}\,.
\label{Ghat}
\eeq
The {\it hat} indicates that $\hat\eta$ is an auxiliary flat metric
while the actual space-time metric is still $G_{MN}$. Now we construct
the spinor representation of the Lorentz twist $\hat\O$ in this
auxiliary flat space. This is obtained from the expression
(\ref{sodd-spinor}) for $\O$ by replacing $\G^i$ by $\hat\G^i$ and
setting $B_{ij}=0$. Furthermore, we can combine the $C'^{(n)}$ into an
$SO(1,9)$ bispinor $\hat{\sC'}$,
\beq
\hat{\sC}\,' =\sum_m \frac{(-1)^m}{m!}\, C\,'^{(m)}_{M_1\cdots M_m}\,
\hat\G^{M_1\cdots M_m}\,.
\label{Cslash}
\eeq
Then the non-trivial $SO(d,d)$ transformations of $C'^{(n)}$ can be
written as
\beq
\t{\hat{\sC}} = \hat\Omega\,\hat{\sC}\,.
\label{Odd-C'}
\eeq

\section{An Application: D-brane Couplings by Clifford Multiplication}

The Abelian theory on a single D-brane couples to the background R-R
potentials by {\it exterior multiplication} ${\cal C}'\wedge e^{\cal
F}$, where ${\cal C}'$ is the pull-back of ${\mathbf C}'$ to the
worldvolume and ${\cal F}$ is the Abelian gauge field strength. In the
non-Abelian case the interaction also involves the scalar products of
the non-Abelian scalars $\Phi^i$ and the R-R potentials. Furthermore,
the factors of $\del X^i/\del\xi^\a$ appearing in the pull-back of
$C'^{(n)}$ in the static gauge are replaced by the gauge covariant
derivatives $D_\a \Phi^i$ \cite{RT,M}. These new interactions are
written in the static gauge and moreover the replacement of the
pull-back by a non-Abelian covariant derivative obscures the geometric
description of the D-brane as an embedded surface.

The construction in the last section allows us to go beyond the static
gauge and write a {\it covariant} and {\it geometric} expression that
includes the new couplings \cite{HM}.  This is achieved by replacing
the exterior product by a {\it Clifford multiplication} defined with
respect to the flat metric $\hat\eta$ as in (\ref{Ghat}). To see this,
note that the degree of a D-brane volume form increases/decreases
under a single T-duality transverse/parallel to the brane. Thus, they
behave very much like the $C'^{(n)}$ in (\ref{td1-C'}) and it makes
sense to construct a volume bispinor $\hat{\sV}$ similar to
(\ref{Cslash}),
\bea
\hat{\sV}\!\!\!&=&\!\!\! \sum _p\hat{\sV}^{(p+1)}=
\sum_p\frac{(-1)^{p+1}}{(p+1)!}\, T_{(p)}
\nonumber\\
&&\times{\rm d} X^{L_1} \wedge\cdots\wedge {\rm d}
X^{L_{p+1}}\,\hat\G_{L_1\cdots L_{p+1}}\,.
\label{Vslash}
\eea
Moreover, the worldvolume gauge fields ${\cal A}_\a(\xi)$ and
transverse scalars $\Phi^I(\xi)$ can be combined into a $1$-form
$A_M(X(\xi))$ such that
\beq
{\cal A}_\a=A_M\del X^M/\del\xi^\a\,,\quad
\Phi^I = A_M\,\hat\eta^{MN}\,a_N^I\,.
\label{Aphi}
\eeq
Here, $\del X^M/\del\xi^\a$ and $a^M_I$ span the tangent and
normal bundles to the worldvolume ($I$ is an orthonormal frame
index). Let $F$ denote the field strength of $A$ and $\mathbf
Y=e^F$. We can also define a bispinor for the generalized D-brane
charges,
\beq
\hat{\sY} = \sum_n \frac{(-1)^n}{n!}\, Y^{(n)}_{N_1\cdots N_n}\,
\hat\G^{N_1\cdots N_n}\,.
\label{Yslash}
\eeq
It is now easy to write a covariant expression for the Dp-brane
couplings to all R-R potentials in terms of a Clifford product,
\beq
I_{WZ}^{(p+1)}=-{\rm Str}\int_{{\cal W}^{(p+1)}}{\rm Tr}
\left(\hat\G_{11}{\overline{\hat{\sV}}}\hat\G_{11}\hat{\sC}'\hat{\sY}\right)\,,
\label{I-Cliff}
\eeq
where ${\rm Str}$ is the symmetrized gauge trace and ${\rm Tr}$ is a
trace over the spinor index. The integral over ${\cal W}^{(p+1)}$
restricts the expression to the Dp-brane worldvolume.

The $\G$-matrix multiplication and tracing can be carried out
and leads to the component form for the generalized WZ action,
including the new interactions. In the static gauge, ($X^\mu=\xi^\mu$,
$X^i=X^i(\xi)$ it reduces to the known form in \cite{RT,M}.
The generalized pull-backs discussed above appear as part of the
restriction of $F_{MN}$ to the D-brane which, using (\ref{Aphi}),
gives
\bea
D_\a\Phi^I&=&\frac{\del X^M}{\del\xi^\a}\,(F_{MN})\hat\eta^{NL}a^I_L
\nonumber\\
&&= \del_\a\Phi^I + \Theta^I_{\a J}\Phi^J + [ {\cal A}_\a, \Phi^I]\,.
\label{nbc}
\eea
The appearance of the normal bundle connection $\Theta^I_{\a J}
=a^I_M\del_\a a^M_J$ here is a reflection of the covariance of the
formalism. However, it does not yet contain the contribution from
curved backgrounds for which it may be necessary to consider the
gravitational couplings of the brane.\par
\vspace{.4cm}
It is a pleasure to thank the organizers of the International
Conference on ``Supersymmetry and Quantum Field Theory'' (Kharkov,
July 25-29, 2000) for the invitation and hospitality.

\end{document}